\documentclass[12pt]{article}
\usepackage{graphicx}
\usepackage{amssymb}
\usepackage{epstopdf}
\newtheorem{theorem}{Theorem}

\newtheorem{observation}[theorem]{Observation}

\def\Exp{{\rm Exp}}
\def\Ln{{\rm Ln}}
\def\0{\emptyset}
\def\e{{\rm e}}
\author{Michael D. Hendy\footnote{
Allan Wilson Centre for Molecular Ecology and Evolution,
Massey University, Palmerston North, New Zealand. {\tt m.hendy@massey.ac.nz}}
\and Sagi Snir\footnote{
{\bf Corresponding author.} Mathematics dept. University of California,
  Berkeley, CA 94720, USA. {\tt ssagi@math.berkeley.edu}}
}

\date{\today}

\title{Hadamard Conjugation for the Kimura 3ST Model:\\ Combinatorial Proof using Pathsets}
\begin{document}
\maketitle
\paragraph{Abstract} In most stochastic models of molecular sequence evolution the probability of each possible pattern of homologous characters at a site is estimated numerically.
However in the case of Kimura's three-substitution-types (K3ST) model, these probabilities can be expressed analytically by Hadamard conjugation as a function of the phylogeny $T$ and the substitution probabilities on each edge of $T$, together with an analytic inverse function.
In this paper we produce a direct proof of these results, using pathset distances which generalise pairwise distances between sequences.
This interpretation allows us to apply Hadamard conjugation to a number of topical problems in the mathematical analysis of sequence evolution.


\section{Introduction}
Hadamard conjugation is an analytic formulation of the relationship between the probabilities of expected site patterns of nucleotides for a set of homologous nucleotide sequences and the parameters of some simple models of sequence evolution on a proposed phylogeny $T$. 
An important application of these relations is to give a theoretical tool to analyse properties of phylogenetic inference, such as the methods of maximum likelihood and maximum parsimony, as well as being a tool for generating simulated data, and determining phylogenetic invariants.
Hadamard conjugation can also be used as directly as for phylogenetic inference, inferring either trees with the Closest Tree algorithm \cite{Hen91,SHSE92} or networks using Spectronet \cite{HLPMH02}.

Hadamard conjugation was first introduced in 1989 \cite{Hen89,HP89} to analyse two-state character sequences evolving under the Neyman model \cite{Ney71}. 
Evans and Speed in 1993 \cite{ES93} noted that Kimura's three substitution types (K3ST) model \cite{Kim81} for 4-state characters could be modelled by the Klein group ${\mathbb Z}_2 \times {\mathbb Z}_2$.
Noting this Sz{\'e}kely et al \cite{SESP93,SSE93} extended the two-state analysis to a more general algebraic theory, where substitutions belonged to an arbitrary Abelian group.
They then applied this to sequences evolving under the K3ST model. 
Current applications of Closest Tree and Spectronet \cite{HLPMH02} are usually applied to the $4-$state K3ST model or its derivatives, the K2ST and Jukes--Cantor models.

A pathset in a phylogenetic tree $T$, is a generalisation of the concept of a paths. 
This approach allows the concept of pairwise distances between sequences to be extended to distances connecting larger sets of taxa. 
It provides properties that can be related to other models, such as the molecular clock hypothesis. 
This has, for example, proved pivotal in allowing a simpler analytic expression of the likelihood function, as developed in~\cite{CHS05}, leading to an algebraic solution for the maximum likelihood points. 
It has also proved useful in identifying phylogenetic invariants~\cite{HP96}, and to the introduction of projected spectra~\cite{WH97} which reduces both the variance in the parameter estimates, and the computational 
complexity of the Closest Tree algorithm~\cite{Hen91}. 
Each of the above examples rely on some  identities between the phylogenetic tree and the probabilities of obtaining sequences evolved under that tree. 
However, these identities were never {\em directly} proved. 
Here we provide for the first time, a direct proof for these identities. Effectively, this is an alternative proof of Hadamard conjugation for the K3ST model, where practical interpretations of the intermediate terms are developed, showing {\em directly} the relationships between the topology of $T$ and the substitution probabilities across its edges. 
This is an important contribution that can serve in the burgeoning
area of algebraic statistics in biology and phylogenetics, in
particular (see e.g.~\cite{AR03,AR04a,AR04b,PS05a,PS05b,SS05} ).  

We model the relationship of the differences of $n$ sequences labeled $1,2, \cdots,n$, from a reference sequence labeled $0$. Because the models are reversible, the choice of reference sequence is arbitrary. 
The topology of $T$ and the model parameters are presented in a sparse matrix $Q_T$ of $2^n$ rows and columns, called the edge-length spectrum. 
The probabilities of each site pattern are presented in a similar sized matrix $P_T$ called the sequence probability spectrum. 
We also define a Hadamard matrix $H_n$ of $2^n$ rows and columns, and show that the matrix products
$$H_nQ_TH_n,\quad H_nP_TH_n,$$
both relate to properties of path-sets. 
We prove the major result by interpreting corresponding components of each entry of these matrices.

In earlier representations \cite{HPS94,SHP98} the Hadamard conjugations for K3ST were presented as a conjugations of  vectors of $4^n$ components by the Hadamard matrices $H_{2n}$ of $4^{n}$ rows and columns.
In the formulation presented here the vectors are replaced by matrices of $2^{n}$ rows and columns, which pre- and post-multiplied by $H_{n}$, a Hadamard matrix of the same order.

\section{Kimura's 3ST model}

Kimura's \cite{Kim81}  three substitution types model (K3ST) specified independent rates, 
$\alpha$, $\beta$ and $\gamma$, for each of three substitution types between the 
RNA or DNA nucleotides. 
Here we will refer to these substitutions as:
\begin{description}
\item[$t_\alpha$: ] the substitutions {\tt A $\leftrightarrow$ G, U(T) $\leftrightarrow$ C} (transitions);
\item[$t_\beta$: ] the substitutions {\tt A $\leftrightarrow$ U(T), G $\leftrightarrow$ C} (transversions type $\beta$);
\item[$t_\gamma$: ] the substitutions {\tt A $\leftrightarrow$ C, U(T) $\leftrightarrow$ G} (transversions type $\gamma$).
\end{description}
By including the identity $t_\epsilon$, we find the set of substitutions 
$${\mathcal T}=\{t_\epsilon,t_\alpha,t_\beta,t_\gamma\}$$
is a group under composition, which acts on the nucleotide set $\{\tt{A,C,G,T(U)}\}$.
\begin{observation}
$({\mathcal T},\circ)$ is isomorphic to the Klein $4-$group, $(\mathbb{Z}_2\times \mathbb{Z}_2,+_2).$ 
\vspace{-10mm} \flushright $\Box$
\end{observation}

Kimura modelled the expected differences between two sequences separated by time $t$. With the three specified rates, the expected numbers of substitutions of each type are therefore
$$q(\alpha)=\alpha t,\quad q(\beta)=\beta t,\quad q(\gamma)=\gamma t.$$
The number of substitutions of each type observed between homologous nucleotides of the two sequences can be used to estimate the probabilities $p(\alpha)$, $p(\beta)$, and $p(\gamma)$ of each type occurring. By setting $\beta=\gamma$, or $\alpha=\beta=\gamma$, this model projects to Kimura's better known two substitution type model \cite{Kim80}, or to the simple Jukes/Cantor model \cite{JC69}.

 Kimura derived expressions for the expected numbers as functions of the probabilities. These are equivalent to the standard expression of the rate matrix $R$ derived from the stochastic matrix $M$, over time $t$,
\begin{equation}\label{eq:A1}
M=\exp(Rt),
\end{equation}
where, with $K=q(\alpha)+q(\beta)+q(\gamma)$ being the total number of substitutions,
$$M=\left[\begin{array}{cccc}p(\epsilon) & p(\alpha) & p(\beta) & p(\gamma) \\p(\alpha) & p(\epsilon) & p(\gamma) & p(\beta) \\p(\beta) & p(\gamma) & p(\epsilon) & p(\alpha) \\p(\gamma) & p(\beta) & p(\alpha) & p(\epsilon)\end{array}\right], \quad
Rt=\left[\begin{array}{cccc}-K & q(\alpha) & q(\beta) & q(\gamma) \\q(\alpha) & -K & q(\gamma) & q(\beta) \\q(\beta) & q(\gamma) & -K & q(\alpha) \\q(\gamma) & q(\beta) & q(\alpha) & -K\end{array}\right].$$

Let $H_2$ be the $4 \times 4$ Hadamard matrix
$$H_2=\left[\begin{array}{rrrr}1 & 1 & 1 & 1 \\1 & -1 & 1 & -1 \\1 & 1 & -1 & -1 \\1 & -1 & -1 & 1\end{array}\right].$$
\begin{observation}
$H_2$ diagonalises both $M$ and $Rt$. In particular
$$H_2^{-1}MH_2=\left[\begin{array}{cccc}1 & 0 & 0 & 0 \\0 & 1-2p(\alpha)-2p(\gamma) & 0 & 0 \\0 & 0 & 1-2p(\beta)-2p(\gamma) & 0 \\0 & 0 & 0 & 1-2p(\alpha)-2p(\beta)\end{array}\right],$$
and
$$H_2^{-1}RtH_2=-2\left[\begin{array}{cccc}0 & 0 & 0 & 0 \\0 & q(\alpha)+q(\gamma) & 0 & 0 \\0 & 0 & q(\beta)+q(\gamma) & 0 \\0 & 0 & 0 & q(\alpha)+q(\beta)\end{array}\right].$$ 
\vspace{-14mm} \flushright $\Box$
\end{observation}
Hence from equation \ref{eq:A1} we find
\begin{eqnarray*}
1-2(p(\alpha)+p(\gamma))=p(\epsilon)-p(\alpha)+p(\beta)-p(\gamma)&=&
\e^{-2(q(\alpha)+q(\gamma)}=\e^{-K-q(\alpha)+q(\beta)-q(\gamma)},\\
1-2(p(\beta)+p(\gamma))=p(\epsilon)+p(\alpha)-p(\beta)-p(\gamma)&=&
\e^{-2(q(\beta)+q(\gamma)}=\e^{-K+q(\alpha)-q(\beta)-q(\gamma)},\\
1-2(p(\alpha)+p(\beta))=p(\epsilon)-p(\alpha)-p(\beta)+p(\gamma)&=&
\e^{-2(q(\alpha)+q(\beta)}=\e^{-K-q(\alpha)-q(\beta)+q(\gamma)},
\end{eqnarray*}
which can be succinctly expressed as
\begin{equation}
\label{eq:A2}
H_1^{-1}PH_1=\Exp(H_1^{-1}QH_1),
\end{equation}
where
$$H_1=\left[\begin{array}{rr}1 & 1 \\1 & -1\end{array}\right],\quad
P=\left[\begin{array}{cc}p(\epsilon) & p(\alpha) \\p(\beta) & p(\gamma)\end{array}\right],\quad
Q=\left[\begin{array}{cc}-K & q(\alpha) \\q(\beta) & q(\gamma)\end{array}\right],$$
and $\Exp$ is the exponential function to each entry of the matrix.
Equation \ref{eq:A2} can be inverted (provided the arguments of $\ln$ are all positive) to give
\begin{equation}\label{eq:A3}
H_1^{-1}QH_1=\Ln(H_1^{-1}PH_1),
\end{equation}
where $\Ln$ is the natural logarithm applied to each component of the matrix.

The invertibility of equations \ref{eq:A2} and \ref{eq:A3} mean that provided the parameters are in the valid ranges, the model could be specified by the three probabilities $p(\alpha)$, $p(\beta)$ and $p(\gamma)$, or by the three parameters $q(\alpha)$, $q(\beta)$ and $q(\gamma)$. Indeed, when we do this, we do not need to rely on a rate/time specification and a Poisson process of substitution.

\section{Substitutions across the edges of a tree}
Let $T$ be a tree (phylogeny) with leaf set $L(T)=\{0,1,\ldots,n\}$, and edge set $E(T)$.
We can postulate three independent Kimura probability parameters $p_e(\alpha)$, $p_e(\beta)$ and $p_e(\gamma)$ for each edge $e \in E(T)$ and a transition matrix
$$M_e=\left[\begin{array}{cccc}p_e(\epsilon) & p_e(\alpha) & p_e(\beta) & p_e(\gamma) \\p_e(\alpha) & p_e(\epsilon) & p_e(\gamma) & p_e(\beta) \\p_e(\beta) & p_e(\gamma) & p_e(\epsilon) & p_e(\alpha) \\p_e(\gamma) & p_e(\beta) & p_e(\alpha) & p_e(\epsilon)\end{array}\right].$$

Suppose we assign nucleotides to each vertex of $T$ according to a model parameterised by these probabilities  for each edge $e \in E(T)$.
The matrices $M_e$ for each $e \in E(T)$ are all diagonalised by $H_2$, and hence commute, so for any subset $W \subseteq E(T)$ of edges we can define
\begin{equation}\label{eq:B1}
M_W=\prod_{e \in W}M_e,
\end{equation}
the transition matrix representing the probabilities of change concatenated across the edges in $W$.

We observe 
\begin{eqnarray*}
H_2^{-1}M_WH_2&=&H_2^{-1}\left(\prod_{e\in W}M_e\right) H_2\\
&=&\prod_{e \in W}H_2^{-1}M_eH_2,
\end{eqnarray*}
is a diagonal matrix whose entries are the products of the corresponding eigenvalues  of the factor matrices $M_e$. 

We can define corresponding $2\times 2$ matrices 
$$P_W=\left[\begin{array}{cc}p_W(\epsilon) & p_W(\alpha) \\p_W(\beta) & p_W(\gamma)\end{array}\right],\quad Q_W=\left[\begin{array}{cc}-K_W & q_W(\alpha) \\q_W(\beta) & q_W(\gamma)\end{array}\right],$$ 
writing $P_e$ for $P_{\{e\}}$, etc.
Hence,  from equation \ref{eq:B1}, the entries $q_W(\alpha)$, $q_W(\beta)$ and $q_W(\gamma)$ of $Q_W$ are linear functions of the logarithms of these eigenvalues, and we find
\begin{observation}
$$Q_W=\sum_{e \in W}Q_e.$$\vspace{-14mm} \flushright $\Box$
\end{observation}
Because of this linearity, we define $q_e(\alpha)$, $q_e(\beta)$ and $q_e(\gamma)$, to be the three {\bf edge-length} parameters, for each edge $e$, and can specify our model by the $3|E(T)|$ independent parameters 
$$q_{e}(\theta)\colon \quad \theta \in \{ \alpha, \beta, \gamma\}; e \in E(T).$$ 

The deletion of an edge $e \in E(T)$ induces two subtrees, whose leaf label sets partition $[n]_0$ into two subsets. We choose that subset $A \in [n]$ ( the subset not containing $0$) to index $e$ as $e_A$.
We incorporate the edge-length parameters into three vectors 
${\bf q}_\alpha$, ${\bf q}_\beta$ and ${\bf q}_\gamma$
indexed by the $2^n$ subsets of $[n]$,
where for $A \subseteq [n]$
$$({\bf q}_\alpha)_A=\cases{q_{e_A}(\alpha)&if $e_A \in E(T), $\cr
-\sum_{e_B \in E(T)}q_{e_B}&if $A = \0,$ \cr
0& else,}$$
with similar structures for ${\bf q}_\beta$ and ${\bf q}_\gamma$.
The entries in these vectors are ordered by the subsets of $[n]$ listed lexicographically: $\0$, $\{1\}$, $\{2\}$, $\{1,2\}$, $\{3\}$, $\{1,3\}$, $\{2,3\}$, $\{1,2,3\}$, $\{4\}$, $\cdots,$ etc.

We will also find it convenient to gather these three vectors into a $2^n \times 2^n$ matrix 
$$Q_T=\left[q_{A,B}\right]_{A,B \subseteq [n]},$$
where
$$q_{A,B}=\cases{q_{e_A}(\alpha)&if $e_A \in E(T), B = \0, $\cr
q_{e_B}(\beta)&if $A=\0,e_B \in E(T), $\cr
q_{e_A}(\gamma)&if $A=B, e_A \in E(T),$\cr
-K_T&if $A = B=\0,$ \cr
0& else,}$$
and
$$K_T=\sum_{e \in E(T)}(q_{e}(\alpha)+q_{e}(\beta)+q_{e}(\gamma))=\sum_{e \in E(T)}K_e.$$
Thus the leading column of $Q_T$ is ${\bf q}_\alpha$, the leading row is ${\bf q}_\beta$, and the leading column is ${\bf q}_\gamma$, all other entries are $0$, apart from the leading entry which is $-K_T$ (hence the sum of all  entries of $Q_T$ is $0$). $Q_T$ is referred to as the {\bf edge length spectrum} for $T$. The positive entries of this spectrum identify the edges of $T$.

\begin{figure}
\begin{center}
$${\bf q}_\alpha=\left[\begin{array}{r}
-K(\alpha)\\q_{1}(\alpha)\\q_{2}(\alpha)\\0\\q_{3}(\alpha)\\q_{13}(\alpha)
\\0\\q_{123}(\alpha)\end{array}\right],
{\bf q}(\beta)=\left[\begin{array}{r}
-K(\beta)\\q_{1}(\beta)\\q_{2}(\beta)\\0\\q_{3}(\beta)\\q_{13}(\beta)
\\0\\q_{123}(\beta)\end{array}\right],
{\bf q}_\gamma=\left[\begin{array}{r}
-K(\gamma)\\q_{1}(\gamma)\\q_{2}(\gamma)\\0\\q_{3}(\gamma)\\q_{13}(\gamma)
\\0\\q_{123}(\gamma)\end{array}\right],
$$
$$Q_T=\left[\begin{array}{rccccccc}
-K&q_{1}(\alpha)&q_{2}(\alpha)&0&q_{3}(\alpha)&q_{13}(\alpha)&0&q_{123}(\alpha)\\
q_{1}(\beta)&q_{1}(\gamma)&.&.&.&.&.&.\\
q_{2}(\beta)&.&q_{2}(\gamma)&.&.&.&.&.\\
0&.&.&0&.&.&.&.\\
q_{3}(\beta)&.&.&.&q_{3}(\gamma)&.&.&.\\
q_{13}(\beta)&.&.&.&.&q_{13}(\gamma)&.&.\\
0&.&.&.&.&.&0&.\\
q_{123}(\beta)&.&.&.&.&.&.&q_{123}(\gamma)
\end{array}\right],$$
\caption{\emph{Example edge length spectra for the tree $T_{13}$ on $n+1=4$ taxa illustrated in figure \ref{fig2}.
Corresponding components of the vectors ${\bf q}_\alpha$, ${\bf q}_\beta$, ${\bf q}_\gamma$, give the three edge lengths parameters for the corresponding edge.
The value ``$0$" value indicates that there is no corresponding edge in $T$.
These vectors are placed in the leading row, column and main diagonal of the matrix $Q$.
This means that for $A, B \subseteq \{1,2,3\}$, $Q_{\0,B}=q_B(\alpha)$, $Q_{A,\0}=q_A(\beta)$, $Q_{A,A}=q_A(\gamma)$, and for all other entries $Q_{A,B}=0$, except the
first entry $Q_{\0,\0}=-K$, where $K=K(\alpha)+K(\beta)+K(\gamma)$.
The entries indicated by ``$.$" are all zero, these are zero for every tree.
The entries indicated by ``$0$" are zero for this tree $T$, but for different trees can be non-zero.
The non-zero entries (in the leading row, column and main diagonal) should each be in the same component, and these identify the edge splits of $T$.
For general trees on $n+1$ taxa, the edge length spectra are vectors and square matrices of order $2^n$. } \label{fig2}}
\end{center}\end{figure}

If we propose a sequence of nucleotides at leaf $0$, then we can generate homologous sequences at each of the other leaves under this model. A common position in each of these sequences is called a site. If in an instance of such sequences, the character states at leaves $0, 1,\ldots,n$ are $\chi(0),\chi(1),\ldots,\chi(n)$, which partitions $[n]$ into the subsets
$$S_{\theta}=\{i \in [n]\colon t_\theta(\chi(0))=\chi(i)\}, \mbox{ for } \theta \in \{\epsilon,\alpha,\beta,\gamma\}.$$
Thus for example $S_\0$ is the set of leaves of $[n]$ with the same state as at $0$.
 We index a site pattern by $(A,B)$, the pair of subsets of $[n]$, where
$$A=S_\alpha \cup S_\gamma, \quad B=S_\beta \cup S_\gamma,$$
noting that the partition can be recovered from $(A,B)$.
In particular 
$$S_\gamma =A\cap B,\,S_\alpha=A-S_\gamma,\,S_\beta=B-S_\gamma,\,S_\epsilon=[n]-(A\cup B).$$
 We will show that the probability $p_{A,B}$ of obtaining the site  pattern $(A,B)$, for each $A,B \in [n]$, is a function of the edge length parameters.

We now define another $2^n \times 2^n$ matrix $P_T$, the {\bf sequence probability spectrum}, with rows and columns indexed by the subsets of $[n]$, where
$$P_T=[p_{A,B}]_{A,B \subseteq [n]},$$
where $p_{AB}$ is the probability of obtaining the site pattern $(A,B)$.

\section{Hadamard matrices and Path-sets}
We define recursively the family $\{H_n\colon n \in {\mathbb Z}\}$, (known as Sylvester matrices), where for $n \ge 2$
$$H_n = H_1 \otimes H_{n-1}=\left[\begin{array}{rr}H_{n-1} & H_{n-1} \\H_{n-1} & -H_{n-1}\end{array}\right]$$
is a symmetric Hadamard matrix of order $2^n$, with $H_1$ and $H_2$ as previously defined. 
It is easily seen that $H_n^{-1}=2^{-n}H_n$.

It is known \cite{SHP98} that if we index the rows and columns of $H_n$ lexicographically by the subsets of $[n]$ that:
\begin{observation}
$$\left[H_n\right]_{A,B}=h(A,B)=(-1)^{|A\cap B|}.$$
\vspace{-14mm} \flushright $\Box$
\end{observation}

Let $\Pi_{i,j}$ be the set of edges in the path in $T$ connecting leaves $i$ and $j$, ($i,j \in \{0,1,\ldots,n\}$) the entries of the transition matrix $M_{\Pi_{i,j}}$ represent the probabilities of observing the corresponding differences between the nucleotides at leaves $i$ and $j$. 
We see further that 
$$M_{\Pi_{i,j}}=\prod_{e_A \in \Pi_{i,j}}M_{e_A}.$$
Because each edge $e_A$ in $E(T)$ separates vertices $i$ from $j$, these edges are precisely those for which $A\cap\{i,j\}$ contains one, but not both elements. Hence we see
$$\Pi_{i,j}=\{e_A \in E(T)\colon h(A,\{i,j\})=-1\}.$$ 
We generalise this, for any $C \subseteq [n]$,  finding it useful to consider the collection of edges
$$\Pi_C=\{e_A\in E(T)\colon h(A,C)=-1\}.$$
\begin{observation}
In \cite{SHP98} it is shown that:\\
for $|C|\equiv 0(\mbox{mod }2)$, $\Pi_C$ is a set of $|C|/2$ edge-disjoint paths, whose endpoints are the leaves in $C$; \\
for $|C|\equiv 1(\mbox{mod }2)$, $\Pi_C$ is a set of $(|C|+1)/2$ edge-disjoint paths, whose endpoints are the leaves in $C\cup\{0\}$. \vspace{-10mm} \flushright $\Box$
\end{observation}
$\Pi_C$ is called a {\bf path-set}. 
In particular $\Pi_{\{i\}}=\Pi_{0,i}$ and $\Pi_{\{i,j\}}=\Pi_{i,j}$ comprise single paths, and $\Pi_\0=\0$. 
We find the set of pathsets is a group (under symmetric difference) isomorphic to ${\mathbb Z}_2^n$.

The sum of edge lengths on a path connecting two leaves can naturally be thought of as the distance between the leaves. We extend this distance concept, for each substitution type $\theta \in \{\alpha,\beta,\gamma\}$ to sets of paths, to define the {\bf path-set distance} 
$$d_{\Pi_C}(\theta)=\sum_{e_A \in \Pi_C}q_A(\theta),$$
so that
\begin{eqnarray}
\sum_{A \subset [n]}h(A,C)q_A(\theta)&=&q_\0+\sum_{e_A \in E(T)}h(A,C)q_A(\theta)\nonumber\\
&=&\sum_{e_A \in E(T)}(-1+h(A,C))q_A(\theta)\nonumber\\
&=&-2\sum_{h(A,C)=-1}q_A(\theta)\nonumber\\
&=&-2d_{\Pi_C}(\theta).
\end{eqnarray}
Suppose each vertex $v$ of $T$ is assigned a character state $\chi(v)$, then for each edge $e=(u,v)\in E(T)$ there is a transformation $t_{\theta_e}$ such that $t_{\theta_e}(\chi(u))=\chi(v)$. We can write $t_{\theta_e}=(\chi(u))^{-1}\chi(v)=\chi(u)\chi(v)$, as the transformations are Boolean.
For the path $\Pi_{i,j}$ connecting leaves $i$ and $j$ we find
$$\prod_{e \in \Pi_{i,j}}t_{\theta_e}=\prod_{e = (u,v) \in E(T)}\chi(u)\chi(v)=\chi(i)\chi(j),$$
as the products at each internal vertex cancel.
Further, for any $C\subseteq [n]$ let $C_0=C\cup\{0\}$ if $|C|$ is odd, $C_0=C$ otherwise, then
\begin{equation}\label{eq:B2}
\prod_{e \in \Pi_C}\theta_e=\prod_{i \in C_0}\chi(i).
\end{equation}
Suppose $\prod_{e \in \Pi_C}\theta_e\in \{\epsilon,\alpha\}$, then the number of factors $\chi(i)\in\{\beta, \gamma\}$ in equation \ref{eq:B2} must be even, hence with $B=S_\beta\cup\gamma$, $h(B,C_0)=h(B,C)=1$. Similarly if $\prod_{e \in \Pi_C}\theta_e\in \{\beta,\gamma\}$, then the number of factors $\chi(i)\in\{\beta, \gamma\}$ in equation \ref{eq:B2} must be odd, so $h(B,C_0)=h(B,C)=-1$. Generalising this we obtain
\begin{observation}
If each vertex $v$ of $T$ is assigned character state $\chi(v)$ with $A=S_\alpha \cup S_\gamma, B=S_\beta\cup\gamma \subseteq [n]$, and for any $C \subseteq [n]$ then for any $C \subseteq \{0,1,\cdots,n\}$ with an even number of elements, let $\chi(C)=\prod_{i \in C}\chi_i$ then
\begin{equation}\label{eq:B3}
\chi(C)\in\{\epsilon,\alpha\}\Leftrightarrow h(B,C)=1,\quad 
\chi(C)\in\{\epsilon,\beta\}\Leftrightarrow h(A,C)=-1.
\end{equation}
\end{observation}

\section{Hadamard Conjugation}
$Q_T$ is the matrix containing the edge weight parameters across $T$. $P_T$ is the matrix of probabilities of patterns at the leaves of $T$. The link between these are the matrix products $H_nP_TH_n$ and $H_nQ_TH_n$ which both relate to pathset properties. 
These enable to state our major result
\begin{theorem}\label{thm:A}
\begin{equation}\label{eq:C1}
P_T=H_n^{-1}(\Exp(H_n^{-1}Q_TH_n))H_n,
\end{equation}
which provided the arguments of the logarithm are positive, is invertible to give
\begin{equation}\label{eq:C2}
Q_T=H_n^{-1}(\Ln(H_nP_TH_n))H_n^{-1},
\end{equation}
\end{theorem}  
{\bf Proof}\\
The proof of this theorem is based on interpreting the corresponding components, 
for $A,B \subseteq [n],$
$$\left[H_nP_TH_n\right]_{A,B}\mbox{ and }\left[H_nQ_TH_n\right]_{A,B}.$$
As the only nonzero entries in $Q_T$ are $Q_{\0,\0}$ and
$Q_{C,\0},Q_{\0,C},Q_{C,C}\colon e_C \in E(T)$, we find
\begin{eqnarray*}
\left[H_nQ_TH_n\right]_{A,B}&=&\sum_{A',B'\subseteq [n]}h(A,A')h(B,B')Q_{A'B'}\\
&=&Q_{\0,\0}+\sum_{e_C \in E(T)}(h(A,C)Q_{C,\0}+h(B,C)Q_{\0,C}+h(A,C)h(B,C)Q_{C,C}\\
&=&\sum_{e_C \in E(T)}((h(A,C)-1)Q_{C,\0}+(h(B,C)-1)Q_{\0,C}+(h(A,C)h(B,C)-1)Q_{C,C})\\
&=&\sum_{e_C \in E(T)}((h(A,C)-1)q_{e_C}(\beta)+(h(B,C)-1)q_{e_C}(\alpha)
+(h(A,C)h(B,C)-1)q_{e_C}(\gamma))\\
&=&-2\sum_{e_C\in \Pi_A}q_{e_C}(\beta)-2\sum_{e_C\in \Pi_B}q_{e_C}(\alpha)
-2\sum_{e_C\in \Pi_A\Delta\Pi_B}q_{e_C}(\gamma)\\
&=&-2\left(d_{\Pi_A}(\beta)+d_{\Pi_B}(\alpha)+d_{\Pi_A\Delta\Pi_B}(\gamma)\right).
\end{eqnarray*}
We can partition $\Pi_A\cup\Pi_B$ into three parts,
$$U=\Pi_A-\Pi_B,\quad V=\Pi_B-\Pi_A, \quad W=\Pi_A\cap \Pi_B,$$
and likewise split the pathset distances into components
$d_U(\theta)=\sum_{e \in U}q_e(\theta)$, etc.
Thus
$$\left[H_nQ_TH_n\right]_{A,B}=-2\left(d_U(\beta)+d_U(\gamma)+d_V(\alpha)+d_V(\gamma)
+d_W(\alpha)+d_W(\beta)\right),$$
and
\begin{equation}
\left[\Exp(H_nQ_TH_n)\right]_{A,B}=\e^{-2(d_U(\beta)+d_U(\gamma))}
\e^{-2(d_V(\alpha)+d_V(\gamma))}
\e^{-2(d_W(\alpha)+d_W(\beta))}.\label{eq:C4}
\end{equation}
Now, by equation \ref{eq:A2},
\begin{eqnarray*}
\e^{-2(d_U(\beta)+d_U(\gamma))}&=&p_U(\epsilon)+p_U(\alpha)-p_U(\beta)-p_U(\gamma),\\
\e^{-2(d_V(\alpha)+d_V(\gamma))}&=&p_V(\epsilon)-p_V(\alpha)+p_V(\beta)-p_V(\gamma),\\
\e^{-2(d_W(\alpha)+d_W(\beta))}&=&p_W(\epsilon)-p_W(\alpha)-p_W(\beta)+p_W(\gamma).
\end{eqnarray*}
Hence equation \ref{eq:C4} becomes
\begin{eqnarray}
\left[\Exp(H_nQ_TH_n)\right]_{A,B}&=&(p_U(\epsilon)+p_U(\alpha)-p_U(\beta)-p_U(\gamma))\nonumber\\
&&\times(p_V(\epsilon)-p_V(\alpha)+p_V(\beta)-p_V(\gamma)\label{eq:C5})\\
&&\times (p_W(\epsilon)-p_W(\alpha)-p_W(\beta)+p_W(\gamma)),\nonumber
\end{eqnarray}
which, when expanded, comprises the sum of $64$ terms of the form
$$\pm p_U(\theta)p_V(\phi)p_W(\psi),\quad \theta,\phi,\psi \in \{\epsilon, \alpha,\beta,\gamma\}.$$
Now consider the joint probability $Pr[\Pi_A\colon\alpha;\Pi_B\colon\beta]$ that the product  of substitutions across the edges of $\Pi_A$ is $t_\alpha$ and the product across the edges of $\Pi_B$ is $t_\beta$.
This event is attained by the combinations of $t_\theta$ across $U$, $t_\phi$ across $V$ and $t_\psi$ across $W$ such that
$$t_\theta t_\psi=t_\alpha \mbox{ and } t_\phi t_\psi=t_\beta,$$
which is attained with $t_\theta=t_\alpha t_\psi$ and $t_\phi=t_\beta t_\psi$, for each $\psi \in \{\epsilon,\alpha,\beta,\gamma\}$, and hence with probability
$$Pr[\Pi_A\colon\alpha;\Pi_B\colon\beta]
=p_U(\alpha)p_V(\beta)p_W(\epsilon)+p_U(\epsilon)p_V(\gamma)p_W(\alpha)
+p_U(\gamma)p_V(\epsilon)p_W(\beta)+p_U(\beta)p_V(\alpha)p_W(\gamma),$$
and these terms each occur with a $+$ sign in equation \ref{eq:C5}. 

Now we see the joint probability that the product of substitutions across $\Pi_A$ is either $\epsilon$ or $\alpha$ and the product across $\Pi_B$ is either $\epsilon$ or $\beta$ is 
$$Pr[\Pi_A\colon\epsilon,\alpha;\Pi_B\colon\epsilon,\beta]=Pr[\Pi_A\colon\epsilon;\Pi_B\colon\alpha]+Pr[\Pi_A\colon\epsilon;\Pi_B\colon\alpha]+Pr[\Pi_A\colon\epsilon;\Pi_B\colon\alpha]+Pr[\Pi_A\colon\epsilon;\Pi_B\colon\alpha],$$
and each summand appears with positive sign in equation \ref{eq:C5}. 
Similar examinations of the terms of equation \ref{eq:C5} gives
\begin{eqnarray}
\left[\Exp(H_nQ_TH_n)\right]_{A,B}
&=&Pr[\Pi_A\colon\epsilon,\alpha;\Pi_B\colon\epsilon,\beta]
+Pr[\Pi_A\colon\beta,\gamma;\Pi_B\colon\alpha,\gamma]\nonumber\\
&&-Pr[\Pi_A\colon\epsilon,\alpha;\Pi_B\colon\alpha,\gamma]
-Pr[\Pi_A\colon\beta,\gamma;\Pi_B\colon\epsilon,\beta].\label{eq:C6}
\end{eqnarray}

Let $A_0$ be the set of endpoints of $\Pi_A$, ($A_0=A$ or $A\cup\{0\}$, whichever is of even order). Then 
$$\prod_{e=(u,v) \in \Pi_A}\chi(u)\chi(v)=\prod_{i \in A_0}\chi(i),$$
as for each internal vertex $w$, $\chi(w)$ occurs twice in the product, which gives the identity. 
Hence
$Pr[\Pi_A\colon\epsilon,\alpha;\Pi_B\colon\epsilon,\beta]$, the joint probability that the product of substitutions across $\Pi_A$ is either $\epsilon$ or $\alpha$ and the product across $\Pi_B$ is either $\epsilon$ or $\beta$, is the joint probability that the product of of states across the leaves of $A_0$ is $\epsilon$ or $\alpha$ and across the leaves of $B_0$ is $\epsilon$ or $\beta$. Thus  by equation \ref{eq:B3}
$$Pr[\Pi_A\colon\epsilon,\alpha;\Pi_B\colon\epsilon,\beta]=
\sum_{A',B'\subseteq [n]\colon h(A,A')=1,h(B,B')=1}P_{A'B'}.$$ 
Similarly we find
\begin{eqnarray*}
Pr[\Pi_A\colon\beta,\gamma;\Pi_B\colon\alpha,\gamma]&=&
\sum_{A',B'\subseteq [n]\colon h(A,A')=-1,h(B,B')=-1}P_{A'B'},\\
Pr[\Pi_A\colon\epsilon,\alpha;\Pi_B\colon\alpha,\gamma]&=&
\sum_{A',B'\subseteq [n]\colon h(A,A')=1,h(B,B')=-1}P_{A'B'},\\
Pr[\Pi_A\colon\beta,\gamma;\Pi_B\colon\epsilon,\beta]&=&
\sum_{A',B'\subseteq [n]\colon h(A,A')=-1,h(B,B')=1}P_{A'B'}.
\end{eqnarray*}
Thus from equation \ref{eq:C6}, combining these probabilities we find
\begin{equation}
\left[\Exp(H_nQ_TH_n)\right]_{A,B}=\sum_{A',B'\subseteq [n]}h(A,A')h(B,B')P_{A'B'},
\end{equation}
giving 
\begin{equation}
\Exp(H_nQ_TH_n)=H_nP_TH_n,
\end{equation}
from which equations \ref{eq:C1} and \ref{eq:C2} follow.
\vspace{-14mm}\flushright$\Box$

\end{document}